\documentclass[twocolumn,showpacs,preprintnumbers,amsmath,amssymb]{revtex4}

\usepackage{times}
\usepackage{natbib}
\usepackage{amssymb,amsbsy,amsmath,amsfonts}
\usepackage{graphicx}% Include figure files
\usepackage{epsf,epsfig,float,latexsym,amsthm,fancyhdr,rotating}
\usepackage{graphics,psfrag,longtable}
\newcommand{\mdif}{(M_n\!-\!\,M_p)_{\textsc{\tiny{QCD}}}}
\newcommand{\mdifs}{\delta M_N^{\textsc{\tiny{QCD}}}}

\newcommand{\mdifqeds}{\delta M_N^{\textsc{\tiny{QED}}}}
\usepackage{color}
\newcommand{\eS}{{\epsilon_S}}
\newcommand{\eP}{{\epsilon_P}}
\newcommand{\res}{{\eS}}
\newcommand{\rep}{{\eP}}
\def\slashA{A \!\!\! \slash}

\begin{document}
\title {Isospin breaking in the nucleon mass and the sensitivity of $\beta$ decays to new physics}
\author{M.~Gonz\'alez-Alonso$^{1,2}$}
\author{J. Martin Camalich$^3$}

\affiliation{$^1$Department of Physics, University of Wisconsin-Madison,
1150 University Ave., Madison, WI, 53706, USA\\
$^2$INFN, Laboratori Nazionali di Frascati, I-00044 Frascati, Italy\\
$^3$Department of Physics and Astronomy, University of Sussex, BN1 9QH, Brighton, UK}
\begin{abstract}
We discuss the consequences of the approximate conservation of the vector and axial currents for the hadronic matrix elements appearing in $\beta$ decay if non-standard interactions are present. In particular the isovector (pseudo)scalar charge $g_{S(P)}$ of the nucleon can be related to the difference (sum) of the nucleon masses in the absence of electromagnetic effects. Using recent determinations of these quantities from phenomenological and lattice QCD studies we obtain the accurate values $g_S=1.02(11)$ and $g_P=349(9)$ in the $\overline{MS}$ scheme at $\mu=2$ GeV. The consequences for searches of non-standard scalar interactions in nuclear $\beta$ decays are studied, finding $\epsilon_S=0.0012(24)$ at 90\%CL, which is significantly more stringent than current LHC bounds and previous low-energy bounds using less precise $g_S$ values. We argue that our results could be rapidly improved with updated computations and the direct calculation of certain ratios in lattice QCD. Finally we discuss the pion-pole enhancement of $g_P$, which makes $\beta$ decays much more sensitive to non-standard pseudoscalar interactions than previously thought. 

\end{abstract}

\maketitle

In pure QCD, the charged $d\rightarrow u$ transitions induce approximately conserved vector and axial currents, 
\begin{eqnarray}
\partial_\mu \left(\bar{u}\gamma^\mu d\right)=-i(m_d-m_u)\bar{u} d, \label{Eq:CVC}\\
\partial_\mu \left(\bar{u}\gamma^\mu\gamma_5 d\right)=i(m_d+m_u)\bar{u}\gamma_5 d, \label{Eq:PCAC}
\end{eqnarray}
with $m_{u,d}$ the respective light-quark masses. These equalities are a particular case of the venerable ``conservation of the vector current'' (CVC) and ``partial conservation of the axial current'' (PCAC) relations, which are derived from global-symmetry considerations~\cite{GellMann:1960np,GellMann:1964tf,Glashow:1967rx} and have become a cornerstone for model-independent approaches to the structure and interactions of hadrons~\cite{Pagels:1974se,Gasser:1983yg}.

A straightforward application of CVC and PCAC concerns the derivation of relations between different hadronic matrix elements of local quark bilinears, or equivalently, relations between the associated form factors. These techniques are customarily used in meson decays to reduce the number of independent form factors required for the description of the hadronic structure of the process (see e.g.~\cite{Cirigliano:2011ny} for kaon decays). 
A well-known application of PCAC to nucleon matrix elements is the Golberger-Treiman relation between the (strong) $\pi N$ coupling and the (weak) nucleon axial coupling $g_A$~\cite{Goldberger:1958vp,Gasser:1987rb,Weinberg:1996kr}.

As shown in the next section, similar relations can be established between the well-known isovector (axial)vector charges $g_{V(A)}$ of the nucleon and their (pseudo)scalar counterparts $g_{S(P)}$ \footnote{Throughout the paper we will use the notation $g_i \equiv g_i (q^2=0)$, where $i=V,A,S,P$.}. The latter are needed to describe nuclear and neutron $\beta$ decays if non-standard (pseudo)scalar interactions are present~\cite{Herczeg:2001vk,Severijns:2006dr,Bhattacharya:2011qm,Cirigliano:2013xha}, and they currently are subject to intensive research mainly through lattice QCD (LQCD) calculations~\cite{Bhattacharya:2013ehc,Green:2012ej}. These investigations are of crucial importance to assess the implications of precise $\beta$ decay measurements to constrain new physics, since the larger $g_{S(P)}$ are, the larger their sensitivity to exotic (pseudo)scalar interactions is.

To make things more interesting, it turns out that the nucleon mass splitting in the absence of electromagnetism $\mdifs\equiv\mdif$ is a necessary input for the calculation of the scalar charge $g_S$. Actually, the isospin corrections to the hadron masses, and in particular to the nucleon mass, are starting to receive much attention. While phenomenological determinations are being revised~\cite{WalkerLoud:2012bg}, different lattice collaborations have embarked on the {\it ab initio} computation of these effects in pure QCD~\cite{Beane:2006fk,Horsley:2012fw,Shanahan:2012wa} or even including QED~\cite{Duncan:1996xy,Duncan:1996be,Blum:2010ym,Aoki:2012st,deDivitiis:2013xla,Borsanyi:2013lga,Basak:2013iw}.

Therefore, chiral symmetry provides a neat connection between these two seemingly unrelated nonperturbative quantities. We show how this can be exploited for translating recent calculations of $\mdifs$ into a precise determination of the scalar charge, which subsequently is used to extract a stringent bound on non-scalar scalar $d\rightarrow u$ transitions from $\beta$-decay data. Inversely, we discuss the implications that recent LQCD calculations of the scalar charge have on the isospin breaking effects in the nucleon mass. Finally, we study the pion pole enhancement of $g_P$ and explore its impact on the $\beta$ decay phenomenology.

\section{Form factors in $\beta$ decay}

The theoretical description of neutron $\beta$ decay within the Standard Model (SM) requires the calculation of the vector and axial hadronic matrix elements, that can be decomposed as follows~\cite{Weinberg:1958ut}
\begin{widetext}
\begin{eqnarray}
&&\langle p(p_p)|\bar{u}\gamma^\mu d|n(p_n)\rangle=\bar{u}_p(p_p)\left[g_V(q^2)\gamma^\mu+\frac{\tilde{g}_{T(V)}(q^2)}{2\overline{M}_N}\sigma^{\mu\nu}q_\nu+\frac{\tilde{g}_S(q^2)}{2\overline{M}_N}q^\mu\right]u_n(p_n),\label{Eq:VecFF}\\
&&\langle p(p_p)|\bar{u}\gamma^\mu\gamma_5 d|n(p_n)\rangle=\bar{u}_p(p_p)\left[g_A(q^2)\gamma^\mu+\frac{\tilde{g}_{T(A)}(q^2)}{2\overline{M}_N}\sigma^{\mu\nu}q_\nu+\frac{\tilde{g}_P(q^2)}{2\overline{M}_N}q^\mu\right]\gamma_5 u_n(p_n),\label{Eq:AxFF}
\end{eqnarray}
\end{widetext}
where $u_{p,n}$ are the proton and neutron spinor amplitudes, $\overline{M}_N$ the average nucleon mass and $q$ the difference between the neutron and the proton momenta, $q=p_n-p_p$. The vector and axial charges, $g_V$ and $g_A$ respectively, are responsible for the leading contributions to the decay rate due to the relatively small energies ($q^2\simeq0$) involved in the process. We have $g_V=1$ up to second order isospin-breaking corrections due to the Ademollo-Gatto theorem~\cite{Ademollo:1964sr}, whereas the axial charge has been accurately measured in $\beta$ decays, $g_A=1.2701(25)\times g_V$~\cite{Beringer:1900zz} assuming that potential new physics contributions can be neglected. Lastly, the sub-leading contributions coming from the so-called \lq\lq induced" form factors $\tilde{g}_i$ are known in the limit of isospin symmetry, a safe approximation at the current level of experimental precision, as discussed in detail in Ref.~\cite{Bhattacharya:2011qm}. The description of nuclear $\beta$ decays requires the introduction of the Fermi and Gamow-Teller nuclear matrix elements that play an analogous role to $g_V$ and $g_A$ in the neutron decay.

If non-standard (pseudo)scalar interactions are present we need to introduce the following matrix elements in the theoretical description
\begin{eqnarray}
&&\langle p(p_p)|\bar{u}\,d|n(p_n)\rangle=g_S(q^2)\bar{u}_p(p_p)u_n(p_n),\label{Eq:ScaFF}\\
&&\langle p(p_p)|\bar{u} \gamma_5 d|n(p_n)\rangle=g_P(q^2)\bar{u}_p(p_p)\gamma_5u_n(p_n),\label{Eq:PseFF}
\end{eqnarray}
whereas a tensor interaction would introduce an additional hadronic matrix element which is not relevant for our discussion~\cite{Bhattacharya:2011qm}. We can see in Eqs.~\eqref{Eq:ScaFF}-\eqref{Eq:PseFF} how the size of the (pseudo)scalar charges modulates the sensitivity of $\beta$ decay experiments to non-standard (pseudo)scalar interactions. It is worth noticing that the pseudoscalar bilinear $\bar{u}_p \gamma_5 u_n$ is itself of order $q/M_N$, what suppresses the contribution of a pseudoscalar interaction~\cite{Jackson:1957zz}. However, we show later how this suppression is partially compensated by an enhanced value of $g_P$. 

\section{Relation with the isospin breaking contribution to the nucleon mass}

Using the CVC result of Eq.~(\ref{Eq:CVC}) in combination with the above-given definitions of the form factors appearing in $\beta$ decay, it is straightforward to derive
\begin{equation}
g_S(q^2)=\frac{\mdifs}{\delta m_q} g_V(q^2) + \frac{q^2/2\overline{M}_N}{\delta m_q} \tilde{g}_S(q^2)~,\label{Eq:Rel0}
\end{equation}
where $\delta m_q=m_d\!-\!m_u$ and once again $\mdifs$ is the difference of neutron and proton masses in pure QCD due to the explicit breaking of isospin symmetry in the up/down quark masses ($m_u \neq m_d$). Notice that the contribution due to electromagnetic effects $\mdifqeds$ is of the same order of magnitude as $\mdifs$, and so the experimental value cannot be used. 

The inclusion of QED in the analysis would modify the CVC relation given in Eq.~\eqref{Eq:CVC} introducing a correction proportional to $\alpha_{\rm e.m.}$. In fact, the extra term in this case, ${e\,\langle p(p_p)|\bar{u}\slashA\,d|n(p_n)\rangle}$, accounts for the (nonperturbative) electromagnetic correction to the nucleon mass difference.

In the limit $q^2\to0$ the expression \eqref{Eq:Rel0} reduces to 
\begin{equation}
g_S=\frac{\mdifs}{\delta m_q}~.
\label{Eq:Rel1}
\end{equation}
up to second order isospin-breaking corrections.
Notice that the renormalization-scale and scheme dependence of the light quark masses and $g_S$ is the opposite, in such a way that the dependence in the scalar charge counterbalances the running of $\epsilon_S$~\cite{Bhattacharya:2011qm,Broadhurst:1994se} rendering the observable quantity $\epsilon_S\,g_S$ scale independent. Throughout this paper we use the $\overline{MS}$ scheme at $\mu=2$ GeV for both the (pseudo)scalar charges and the light quark masses.

Likewise, using PCAC in Eq.~(\ref{Eq:PCAC}) one can obtain the following relation
\begin{equation}
g_P(q^2)=\frac{\overline{M}_N}{\overline{m}_q} g_A(q^2) + \frac{q^2/2\overline{M}_N}{\left( 2\overline{m}_q\right)} \tilde{g}_P(q^2)~,
\end{equation}
with $\overline{m}_q$ is the average light-quark mass and where we have dropped the \lq\lq QCD" subindex in the average nucleon mass, since, in this case, all isospin breaking contributions represent small corrections and we can just use the experimental value of the nucleon masses. At zero momentum transfer $q^2\to0$ this expression reduces to 
\begin{equation}
g_P=\frac{\overline{M}_N}{\overline{m}_q} g_A, \label{Eq:Rel2}
\end{equation}     
where considerations similar to those for the scalar charge regarding the renormalization apply. 

Before discussing the phenomenological applications of these relations, let us mention that they have been discussed previously in the context of electric dipole moments \cite{Engel:2013lsa,Anselm:1985cf,Ellis:2008zy}, where one encounters isospin rotations of the hadronic matrix elements of $\beta$ decay.

\section{Numerical analysis}

\begin{table}[t]
\centering
\caption{Summary of results for the isospin breaking contribution to the nucleon mass difference in QCD $\mdifs$. For comparison, we also show the results obtained using Eq.~\eqref{Eq:Rel1} (CVC) with the quark mass difference reported by FLAG~\cite{Colangelo:2010et} and the LQCD calculations of $g_S$ by the LHPC~\cite{Green:2012ej} and PNDME~\cite{Bhattacharya:2013ehc} collaborations.
\label{Table:mnmp}}
\begin{ruledtabular}
\begin{tabular}{ccc}
Type&Label&$\mdifs$ [MeV]\\
\hline
Pheno		&	WLCM~\cite{WalkerLoud:2012bg}			&	2.59(47)\\
LQCD 		&	NPLQCD~\cite{Beane:2006fk}				&2.26(57)(42)\\
LQCD 		&	QCDSF-UKQCD~\cite{Horsley:2012fw}	&3.13(15)(16)(53)\\
LQCD 		&	STY~\cite{Shanahan:2012wa}				&2.9(4)\\
LQCD 		&	RM123~\cite{deDivitiis:2013xla}			&2.9(6)(2)\\
LQCD 		&	BMW~\cite{Borsanyi:2013lga}				&2.28(25)(7)(9)\\
\hline
LQCD+CVC	&	LHPC~\cite{Green:2012ej}				&2.72(83)\\
LQCD+CVC	&	PNDME~\cite{Bhattacharya:2013ehc}	&1.66(62)
\end{tabular}
\end{ruledtabular}
\end{table}

The determination of the scalar charge through the above-derived relation requires the knowledge of the light-quark mass difference $\delta m_q$, and its contribution to the nucleon mass splitting, $\mdifs$.

Interestingly enough, these quantities and, more broadly, the understanding of the interplay between isospin breaking due to the quark masses and electromagnetic effects, have become a topic of very intensive research. On the one hand, the classic phenomenological determination of the QED contributions to $\delta M_N\!=\!M_n\!-\!M_p$ using the Cottingham's sum rule~\cite{Cottingham:1963zz,Gasser:1974wd} has been recently updated~\cite{WalkerLoud:2012bg}, leading to the value $\mdifqeds=-1.30(3)(47)$ MeV. In addition, the uncertain knowledge on the isovector magnetic polarizability of the nucleon has been singled out as the major source of uncertainty. Experimental prospects for measuring this observable~\cite{Lundin:2002jy,Weller:2009zza} will presumably improve the accuracy of the phenomenological extraction of $\mdifqeds$ in the future~\cite{WalkerLoud:2012bg}. We list in the first row of Table~\ref{Table:mnmp} the result on $\mdifs$ obtained using the current determinations in combination with the experimentally measured $\delta M_N=1.2933322(4)$ MeV~\cite{Beringer:1900zz}.  

\begin{figure}[t]
\vspace{0.5cm}
\epsfig{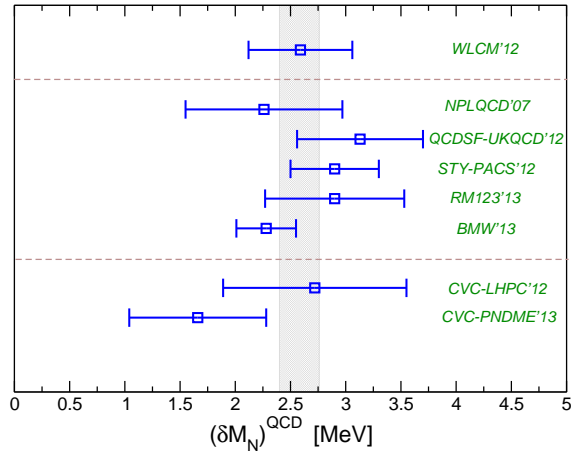} 
\caption{(Color on-line) 
Graphic representation of the different results for the isospin breaking contribution to the nucleon mass difference in QCD $\mdifs$ summarized in Table~\ref{Table:mnmp}, along with the average performed in this work (gray shaded band). 
\label{fig_mnmp}}
\end{figure}

On the other hand, LQCD collaborations are starting to implement isospin breaking effects~\cite{Beane:2006fk,Horsley:2012fw,Shanahan:2012wa} or even directly simulating QED together with QCD~\cite{Duncan:1996xy,Duncan:1996be,Blum:2010ym,Aoki:2012st,deDivitiis:2013xla,Borsanyi:2013lga,Basak:2013iw} (for a recent review see ref.~\cite{Portelli:2013jla}). In addition to the calculation by the NPLQCD collaboration in 2006~\cite{Beane:2006fk}, we consider in our analysis five new determinations reported in the last two years by QCDSF-UKQCD,~\cite{Horsley:2012fw}, Shanahan {\it et al.}~\cite{Shanahan:2012wa}, RM123~\cite{deDivitiis:2013xla} ($N_f=2$) and BMW~\cite{Borsanyi:2013lga} collaborations. Their results are listed in Table~\ref{Table:mnmp} and shown in Fig.~\ref{fig_mnmp}, where the good agreement among them is clearly seen. The additional determination by Blum {\it et al.}~\cite{Blum:2010ym} yielded $\mdifs=2.51(14)$ MeV (statistical error only), but given the absence of an estimate of systematic errors it has not been included in our subsequent numerical analysis. We do not consider either the results obtained in quenched LQCD by Duncan {\it al.} in their seminal work~\cite{Duncan:1996be}.

We combine in quadrature the statistical and systematic errors quoted for each determination, and we obtain from these six determinations the following weighted average
\begin{equation}
[\mdifs]_{av}=2.58(18)~{\rm MeV}~, \label{Eq:Resultmnmp}
\end{equation}
with $\chi^2/{\rm d.o.f.}=0.64$.

The other ingredient needed to calculate the scalar form factor using Eq.~\eqref{Eq:Rel1} is the difference of up and down quark masses. This quantity is usually not quoted by the averaging collaborations, but it can be calculated from the $m_{u,d}$ averages assuming that the correlation between them can be neglected. In this way we obtain $\delta m_q=2.52(19)$ MeV (FLAG)~\cite{Colangelo:2010et} and $\delta m_q=2.55(25)$ MeV (PDG)~\cite{Beringer:1900zz}, where the errors have been added in quadrature.

Using the result of the FLAG collaboration for the quark mass difference, and the above-given average for the nucleon mass splitting in pure QCD, the CVC relation given in Eq.~\eqref{Eq:Rel1} yields
\begin{equation}
g_S=1.02(8)_{\delta m_q}(7)_{\delta M_N}=1.02(11),\label{Eq:Resultgs}
\end{equation}
where the errors stemming from $\delta m_q$ and $\mdifs$ are first shown separately and then added in quadrature. Notice how the former gives a sizable contribution to the total uncertainty. 

It is worthwhile stressing that our result for $g_S$ has been obtained ignoring possible correlations between numerator and denominator of Eq.~(\ref{Eq:Rel1}), as we did when calculating the quark mass difference. These assumptions would be unnecessary in a direct LQCD calculation of the ratio $\mdifs / \delta m_q$, which should be fairly simple to implement in future LQCD analyses. 

This determination of $g_S$ is significantly more precise than direct LQCD calculations available in the literature. The LHPC finds $g_S=1.08(32)$~\cite{Green:2012ej}, whereas the PNDME collaboration has recently published the result $g_S=0.66(24)$~\cite{Bhattacharya:2013ehc}, which supersedes their original preliminary estimate $g_S=0.8(4)$~\cite{Bhattacharya:2011qm}. Inversely, these calculations provide independent determinations of $\mdifs$. By using again Eq.~(\ref{Eq:Rel1}), we obtain the two points shown in the lower part of Table~\ref{Table:mnmp} and Fig.~\ref{fig_mnmp}, corresponding to the value of $g_S$ obtained by the LHPC and the last one reported by the PNDME collaboration. We see that these determinations are starting to have an accuracy close to the direct calculations of $\mdifs$ and that the result from the PNDME collaboration marginally disagrees with the average in Eq.~(\ref{Eq:Resultmnmp}).

Likewise, the application of PCAC through Eq.~\eqref{Eq:Rel2} yields the following result for the pseudoscalar charge
\begin{equation}
g_P=349(9)~,\label{Eq:ResultgP}
\end{equation}
where the error is entirely dominated by the error in $\overline{m}_q=3.42(9)$ MeV, value that is taken from the $N_f=2+1$ FLAG average~\cite{Colangelo:2010et}. Notice the large enhancement experienced by this form factor that diverges in the chiral limit. This effect is due to the pole of a charged pion present in the coupling of a pseudoscalar field to the $d\,u$ vertex in QCD at low energies. In fact, this result is equivalent to the Goldberger-Treiman relation in which the pseudoscalar current serves as an interpolator of the pion field and $g_P(q^2)$ is expressed as a function with a pole at $q^2=M_\pi^2$, whose residue is defined as the strong pion-nucleon coupling (see e.g. Ref.~\cite{Gasser:1987rb,Weinberg:1996kr} for details).
 
\section{Implications for New Physics searches in $\beta$ decays}
Given the V-A structure of the weak interaction, the (pseudo)scalar hadronic matrix elements of Eq.~\eqref{Eq:ScaFF}-\eqref{Eq:PseFF} do not appear in the SM description of $\beta$ decays. However the contribution due to new physics, like the coupling of a heavy charged scalar to first generation fermions, would require the calculation of these matrix elements. Such non-standard interactions can be described by a low-energy effective Lagrangian for semi-leptonic $d\to u e \nu$ transitions~\cite{Cirigliano:2009wk,Cirigliano:2012ab}, where the scalar and pseudoscalar interactions are described by the following terms:
\begin{eqnarray}
{\cal L}_{d\to u e \nu}  &=& {\cal L}^{SM}_{d\to u e \nu}
- \frac{G_F V_{ud}}{\sqrt{2}} \  
\Big[ \ 
\eS  \  \  \bar{e}  (1 - \gamma_5) \nu_e  \cdot  \bar{u} d 
\nonumber \\
 &-& \eP  \  \   \bar{e}  (1 - \gamma_5) \nu_e  \cdot  \bar{u} \gamma_5 d 
\Big]+{\rm h.c.}
\label{eq:Leff}
\end{eqnarray}
Here $\nu_e,e,u,d$ denote the electron neutrino, electron, up- and down-quark  mass eigenfields, whereas $\eS$ and $\eP$ are the Wilson coefficients generated by some unspecified non-standard dynamics. Notice we do not consider interactions involving right-handed neutrinos, that in any case would not interfere with the SM contributions, making their impact on the observable substantially smaller~\cite{Cirigliano:2012ab}. 
Moreover, and for the sake of simplicity we will assume in this work that the Wilson coefficients $\epsilon_{S,P}$ are real, corresponding to CP-conserving interactions.

\subsection{Scalar interaction}

The most stringent limits on non-standard scalar interactions obtained from $\beta$ decays arise from the contribution of the Fierz interference term to the $\mathcal{F}t$-values of super-allowed pure Fermi transitions \cite{Hardy:2008gy}, namely
\begin{equation}
b_F = -2\,g_S\, \res= -0.0022(43) ~~~~~~\mbox{(90\% CL)}~.%,
\label{eq:bF}
\end{equation}
Alternative bounds on scalar interactions can be also obtained from the measurement of the $\beta\nu$ angular correlation $a$ in pure Fermi transitions, also through the Fierz term. Although several on-going and planned experiments will improve the current measurements of $a$, it seems unlikely that they will be able to improve the bound given in Eq.~\eqref{eq:bF} in the near future~\cite{Gonzalez-Alonso:2013uqa}. On the other hand, the Fierz term in neutron $\beta$ decay is also sensitive to scalar interactions, although the level of precision required to compete with the bounds from nuclear decays looks also quite challenging, at least for the current generation of experiments~\cite{Bhattacharya:2011qm}. 

Given the experimental measurement of Eq.~\eqref{eq:bF} and the value of the scalar form factor derived in the previous section, we can determine the current bound on $\res$ from $\beta$ decays. Following Refs.~\cite{Bhattacharya:2011qm,Cirigliano:2012ab,Gonzalez-Alonso:2013uqa} we calculate the confidence interval on $\res$ using the so-called R-Fit method~\cite{Hocker:2001xe}. In this scheme the theoretical likelihoods do not contribute to the $\chi^2$ of the fit and the corresponding QCD parameters take values within certain \lq\lq allowed ranges". In our case, this means that $g_S$ is restricted to remain inside a given interval, namely $0.91 \leq  g_S  \leq 1.13$ from Eq.~\eqref{Eq:Resultgs}. Notice that all values inside this range are treated on an equal footing, whereas values outside the interval are not permitted. Also note that the bound on $\res$ depends only on the lower limit of the scalar form factor, as long as $b_F$ is compatible with zero at $1\sigma$.

\begin{figure}[t]
\vspace{0.5cm}
\epsfig{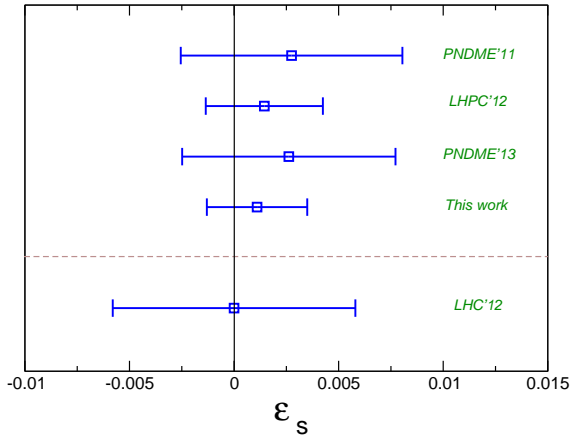} 
\caption{(Color on-line) Bounds on $\eS$ at 90\% CL from the measurement of $b_F$ in super-allowed pure Fermi transitions \cite{Hardy:2008gy}, Eq.~\eqref{eq:bF} using different values for the scalar form factor $g_S$~\cite{Bhattacharya:2011qm,Green:2012ej,Bhattacharya:2013ehc}. For comparison we show also the bound obtained from the analysis of LHC data carried out in Ref.~\cite{Gonzalez-Alonso:2013uqa}.}
\label{fig:es}
\end{figure}

In this way we obtain the following limit on CP-conserving scalar interactions
\begin{equation}
\res  =  0.0012(23)~~~~~\mbox{(90\% CL)}~,
\label{eq:esresult}
\end{equation}
which, as it is shown in Fig.~\ref{fig:es}, improves significantly the bound obtained in Ref.~\cite{Bhattacharya:2011qm} using $g_S=0.8(4)$. Fig.~\ref{fig:es} shows also the $\res$ bound that we obtain using the later LQCD calculations of $g_S$ done by the LPHC~\cite{Green:2012ej} and PNDME~\cite{Bhattacharya:2013ehc} collaborations. It is worth mentioning that if we abandoned the R-fit scheme and treated $g_S$ as a normally distributed variable we would obtain $\res  =  0.0011(21)$ at 90\% CL, in good agreement with the R-fit result of Eq.~\eqref{eq:esresult}.

The LHC searches can also be used to set bounds on $\eS$ and $\eP$. This can be done in a model-independent way if the new degrees of freedom that produce the effective scalar interaction in $\beta$ decays are also too heavy to be produced on-shell at the LHC, since in that case it is possible to study collider observables using a high-energy $SU(2)_L\times U(1)_Y$-invariant effective theory that can be connected to the low-energy effective theory of Eq.~\eqref{eq:Leff}. 

In Fig.~\ref{fig:es} we show the most stringent bound on $\eS$ from LHC searches, obtained in Ref.~\cite{Gonzalez-Alonso:2013uqa} studying the channel $pp\to e+ {\rm MET}+X$, where MET stands for missing transverse energy. More specifically, a CMS search with 20 fb$^{-1}$ of data recorded at $\sqrt{s}=$ 8 TeV~\cite{ENcms20fb}, was used to obtain $|\epsilon_{S,P}|< 5.8\times 10^{-3}$ at 90\% CL~. Notice that these searches are equally sensitive to scalar and pseudoscalar interactions.

\subsection{Pseudoscalar interaction}

In the study of the effect of non-standard interactions in nuclear and neutron $\beta$ decays, it is common lore that the pseudoscalar terms can be safely neglected in the analysis because the associated hadronic bilinear $\bar{u}_p \gamma_5 u_n$ is of order $q/M_N$, which represents a suppression of order $\sim 10^3$. However, we showed in the previous section how the application of PCAC yields $g_P=348(11)$ for the pseudoscalar charge, reducing considerably the suppression from the pseudoscalar bilinear. This result means that, modulo numerical factors of order one, $\beta$ decays with a non-zero Gamow-Teller component, are as sensitive to pseudoscalar interactions as they are to scalar and tensor couplings. 

As a representative example we show here the leading contribution of a non-zero pseudoscalar interaction to the electron energy spectrum in the $\beta$ decay of an unpolarized neutron
\begin{eqnarray}
\hspace{-0.6cm}\frac{d\Gamma}{dE_e} \!&=&\! \frac{G_F^2  \,  |V_{ud}|^2 (1 + 3\lambda^2)}{2\pi^3} p_e E_e (E_0\!-\!E_e)^2 (1+\!\delta_P)\,,
\end{eqnarray}
where $E_e$ and $p_e$ denote the electron energy and the modulus of the three-momentum, $E_0=\delta M_N-(\delta M_N^2-m_e^2)/(2M_N)$ is the electron endpoint energy and $m_e$ is the electron mass. (For sub-leading SM effects see Refs.~\cite{Bhattacharya:2011qm,Ando:2004rk}). The non-standard contribution $\delta_P$ coming from a non-zero effective coupling $\eP$ is given by
\begin{eqnarray}
\delta_P &=& -\frac{\lambda}{1 + 3\lambda^2}~ g_P\, \rep\,\frac{E_0-E_e}{M_N} \,\frac{m_e}{E_e}~,
\end{eqnarray}
where the factor $(E_0-E_e)/M_N$ represents the above-discussed suppression from the pseudoscalar bilinear.

For the sake of comparison, we show now the (well-known) correction stemming from a scalar coupling $\eS$
\begin{eqnarray}
\delta_S &=& \frac{2}{1 + 3\lambda^2}~ g_S\, \res \,\frac{m_e}{E_e}~.
\end{eqnarray}
In the best case $E_e=m_e$ we have then $\delta_P \approx -0.06\,\rep$ and $\delta_S \approx 0.36\,\res$, and the pseudoscalar contribution is only a factor 6 smaller than the scalar one.

This is certainly an interesting result that deserves more detailed studies, in particular related to the sensitivity of current and future $\beta$ decay measurements to $\eP$. We hasten to add that the this coupling happens to be very strongly constrained by the helicity-suppressed ratio $R_\pi \equiv \Gamma(\pi \to e \nu [\gamma])/\Gamma(\pi \to \mu \nu [\gamma])$~\cite{Britton:1992pg,Czapek:1993kc,Cirigliano:2007xi}. It should be noticed, however, that there are some possible loopholes in the bound from leptonic pion decays, like the cancellation of effects between the electron and muon channel in $R_\pi$ due to an $\eP$ coupling proportional to the lepton masses, or cancellations between linear and quadratic terms originated from flavor non-diagonal contributions or interactions with right-handed neutrinos~\cite{Bhattacharya:2011qm,Herczeg:1994ur,Herczeg:2001vk}. In this sense, the extraction of bounds on the effective pseudoscalar coupling $\eP$ from $\beta$ decays paves the road for studying these scenarios.

\section{Conclusions}

In summary, we have discussed the application of the CVC and PCAC relations of QCD to connect different form factors describing $\beta$-decays in the SM and beyond. 

On one hand, we found that CVC relates the scalar charge $g_S$ to the vector charge $g_V\!\!\approx\!\!1$, the isospin breaking of the nucleon mass in pure QCD and the mass splitting $m_d-m_u$. Using a set of recent phenomenological and LQCD determinations of these quantities we obtained a value for $g_S$ with an uncertainty much smaller than the one reported by direct LQCD calculations of this form factor, cf. Eq.~(\ref{Eq:Resultgs}). In turn, we discussed the consequences of this novel determination to the bounds set on non-standard scalar interactions from $\beta$-decays, finding the limit on $\res$ given in Eq.~(\ref{eq:esresult}), which is much stronger than in previous analyses of $\beta$-decays or than those currently obtained from the LHC. 

On the other hand, PCAC relates the pseudoscalar charge $g_P$ to the axial-vector one $g_A$, and the sums of the nucleon and quark masses. We found that $g_P$ is enhanced by the pion pole, counterbalancing the suppression that the contribution of $\epsilon_P$ to the $\beta$ decay suffers from the pseudoscalar quark bilinear. This opens the possibility to investigate scenarios with exotic pseudoscalar contributions that cannot be probed with leptonic pion decays.
%for which leptonic pion decays are insensitive to.

\section*{Acknowledgments}
%%%%%%%%%%%%%%%%%%%%%%ACKNOWLEDGMENTS%%%%%%%%%%%%%%%%%%%%%%%%%%%%%
%
This work was supported in part by the U.S. DOE contract DE-FG02-08ER41531, the Wisconsin Alumni Research Foundation, the Science Technology and Facilities Council (STFC) under grant ST/J000477/1, the Spanish Government and FEDER funds under contract FIS2011-28853-C02-01 and the Fundaci\'on S\'eneca project 11871/PI/09.

%%%%%%%%%%%%%%%%%%%%%%%BIBLIOGRAPHY%%%%%%%%%%%%%%%%%%%%%%%%%%%%%%%%%
%

%
%%%%%%%%%%%%%%%%%%%%%%%BIBLIOGRAPHY%%%%%%%%%%%%%%%%%%%%%%%%%%%%%%%%%

\end{document}